# A Symmetry and Graph Regularized Nonnegative Matrix Factorization Model for Community Detection

Zhigang Liu, and Xin Luo, *Senior Member, IEEE*

*Abstract*—Community is a fundamental and critical characteristic of a Large-scale Undirected Network (LUN) like a social network, making community detection a vital yet thorny issue in LUN representation learning. Owing to its good scalability and interpretability, a Symmetric and Nonnegative Matrix Factorization (SNMF) model is commonly used to tackle this issue. However, it adopts a unique Latent Factor (LF) matrix for precisely representing an LUN's symmetry, which leads to a reduced LF space that impairs its representational learning ability. Motivated by this discovery, this study proposes a <u>S</u>ymmetry and <u>G</u>raph-regularized <u>N</u>onnegative <u>M</u>atrix <u>F</u>actorization (SGNMF) method that adopts three-fold ideas: a) leveraging multiple LF matrices to represent an LUN, thereby enhancing its representation learning ability; b) introducing a symmetry regularization term that implies the equality constraint between its multiple LF matrices, thereby illustrating an LUN's symmetry; and c) incorporating graph regularization into its learning objective, thereby illustrating an LUN's intrinsic geometry. A theoretical proof is provided to ensure SGNMF's convergence on an LUN. Extensive experiment results on ten LUNs from real applications demonstrate that our proposed SGNMF-based community detector significantly outperforms several baseline and state-of-the-art models in achieving highly-accurate results for community detection.

## I. INTRODUCTION

Networks are ubiquitous in our life in the age of Internet. Numerous entities in a real system along with their interactions come into being a Large-scale Undirected Network (LUN), e.g., a social network [1], [2], [4], biological network [3], and wireless sensor network [4]. To extract useful knowledge from LUN, its representation learning has become an important research direction and drawn wide attention from different disciplines [9]-[11], where community detection is an essential issue.

Commonly, a community can be considered as its sub-graph that consists of a set of nodes tightly connected with each other through direct or indirect connections [5], [7]. For instance, a person commonly communicates much more frequently with people in his/her own social circle than those in other circles. Thus, his/her own circle can be considered as a community in a social network. Recent studies [5]-[8] show that communities are ubiquitous in a real network, and play a crucial part in revealing the mechanism of organization and operation of LUN. Based on accurately detected communities, various network analysis tasks are facilitated, e.g., edge prediction [12], social recommendation [13], and opinion behavior analysis [14].

During the last decade, community detection has attracted great attention from researchers [11], [15], leading to a pyramid of detection models [15]-[21]. Among them, a Non-negative Matrix Factorization (NMF) model [22]-[31] has proven to be of high interpretability and scalability. It is frequently applied to the task of community detection on LUNs. Given an LUN, an NMF-based community detector works by building a low-rank approximation to its adjacency matrix with two nonnegative Latent Factor (LF) matrices. The LF matrices are found by minimizing a cost function defined on the LUN's adjacency matrix and its low-rank approximation, thereby accurately representing the topology of the target LUN. The achieved LFs can be either considered as the soft-threshold to identify the community of a specific entity [33], or as the input of a hierarchical community detector [32].

Zhang *et al*. [22] propose a standard NMF-based community detector, which performs approximate factorization on a target LUN's diffusion kernel-based similarity matrix. It is able to identify fuzzy communities in LUN. A Bayesian NMF model is proposed by Psorakis *et al*. [24] to achieve soft partitioning of LUN efficiently, thereby discovering overlapping communities from LUN. Ye *et al*. [26] develop an autoencoder-based NMF model, which is able to capture complex hierarchical mappings between the community membership space and the target network, thereby accurately identifying communities from a complex network. Leng *et al*. [23] present a graph regularized $L_p$ smooth NMF model for data representation. It considers the intrinsic geometry of

✧ Z. Liu is with the School of Computer Science and Technology, Chongqing University of Posts and Telecommunications, Chongqing 400065, China, and with the Chongqing Institute of Green and Intelligent Technology, Chinese Academy of Sciences, Chongqing 400714, China (email: liuzhigangx@gmail.com).
✧ X. Luo is with the College of Computer and Information Science, Southwest University, Chongqing 400715, China. (luoxin21@gmail.com)

target data to produce smooth and stable solution. By incorporates the prior information into the factorization process, a semi-supervised joint NMF algorithm is presented by Ma *et al*. [27] to improve the community detection accuracy in a multi-layer network.

The above-mentioned approaches, despite of their efficiency in community detection, fail to correctly represent LUN's symmetry that is its intrinsic characteristic. To address this problem, a Symmetric and Nonnegative Matrix Factorization (SNMF) model is proposed, which mostly works by adopting a unique LF matrix to learn the low-rank approximation to an LUN's adjacency matrix, thereby correctly representing its symmetry. As discussed in [32], SNMF can well be used to perform both overlapping and non-overlapping community detection. Ding *et al*. [33] prove that SNMF is equivalent to the Laplacian-based spectral clustering and kernel *k*-means. Yang *et al*. [10] present a unified interpretation to SNMF-based community detectors, and propose a unified semi-supervised framework that combines both prior information and network topology to improve detection accuracy. Shi *et al*. [36] propose a pairwisely constrained SNMF model that leverages pairwise constraints generated from partial ground-truth information into its loss function to enhance detection accuracy. Ye *et al*. [28] present a homophily preserving SNMF approach that combines the link topology and node homophily (i.e., the node proximity revealed by specific similarity measurements) of a network, thereby better describing community structures. Luo *et al*. [39] propose to control the scaling-factor in a nonnegative multiplicative update (NMU) rule for an SNMF-based model, resulting in several highly-accurate community detectors.

According to state-of-the-arts [10], [32]-[40], nonnegativity and symmetry constraints enable an SNMF-based community detector's high efficiency and great interpretability. However, existing SNMF models commonly adopt a single LF matrix only for guaranteeing its rigorous symmetry, which is actually a so strong constraint that restricts its representation learning ability [41]-[47]. Motivated by this discovery, we present and answer the following research question via this study:

**Is it possible to build an NMF-based community detector that represents LUN's symmetry with multiple LF matrices, thereby achieving highly accurate detection results?**

To answer this critical question, we present a <u>S</u>ymmetry and <u>G</u>raph-regularized <u>N</u>onnegative <u>M</u>atrix <u>F</u>actorization (SGNMF) model relying on three ideas: a) Adopting multiple LF matrices to represent LUN, thereby enhancing its representation ability; b) Introducing an equality regularization term into its learning objective, thereby making the resultant model well-aware of the target LUN's symmetry; and c) Incorporating a graph regularization term into its learning objective for illustrating the target LUN's intrinsic geometry. This study aims to make the following contributions:

a) **An SGNMF model.** It models an LUN's symmetry by leveraging a symmetry regularization term that implies equality of its multiple LF matrices into its learning objective, thereby expanding LF space to enhance its representation ability. It further models a target LUN's intrinsic geometry by incorporating the graph regularization into its learning model, thereby ensuring its local invariance. Based on such designs, it achieves high representation ability to LUN.
b) **Rigorous convergence analysis of SGNMF.** This analysis is implemented through two separate steps, i.e., proving the learning objective's non-increasing tendency by an auxiliary function-based scheme, and proving the LF learning sequence's convergence to a stationary point of the learning objective by analyzing its Karush-Kuhn-Tucker (KKT) conditions.

Empirical studies on ten real LUNs from industrial applications show the proposed SGNMF model's significant performance gain in community detection accuracy compared with several baseline and state-of-the-art methods. Hence, SGNMF's excellent representation learning ability to an LUN is clearly demonstrated.

II. PROBLEM STATEMENT

*A. Problem Formulation*

A community detector models an LUN with a graph $G=(V, E)$ as $V$ represents a set of $n$ nodes and $E$ represents a set of $m$ edges. Recall the definition of a community detector [15].

***Definition* 1.** *A Community Detector*. Given an LUN $G$ and corresponding adjacency matrix $A^{n \times n}$ whose element $a_{ij}$ is one if $e_{ij} \in E$ and zero otherwise, $\forall v_i \in V$, a community detector aims to find its proper affiliation to its closely related nodes.

Given $G$ and $A$, an NMF-based community detector assumes that there are $K$ communities in $G$ with $K$ being known as prior information, and then builds a rank-$K$ approximation to $A$ as $\hat{A}=XY^T$ (*s.t.* $X, Y \geq 0$). Afterwards, matrix $Y$ can be taken as the membership indicator [39]: $\forall j \in \{1 \sim n\}$, $k \in \{1 \sim K\}$, $y_{jk}$ indicates the probability that node $v_j$ can be assigned to community $C_k$. The above process is formulated by:

$$\forall v_j \in V : \forall v_j \in C_k, \text{ if } y_{jk} = \max\{y_{jl} | l \in \{1 \sim K\}\}. \tag{1}$$

*B. An NMF-based Community Detector*

Considering the generative process of a network revealed in [24], an arbitrary entry $a_{ij}$ in matrix $A$ denotes the interaction probability between nodes $v_i$ and $v_j$, which is an observed variable. In the context of community detection, it is assumed that there are $K$ node clusters, i.e., communities, in a target network which affect the element $a_{ij}$. Hence, we can define an allocation matrix $Y^{n \times K}$ to describe the probability distribution that $n$ nodes belong to $K$ communities as a latent variable and build it by a widely used NMF framework [48]-[55].

Given an LUN, an NMF-based method aims to figure out a low-rank approximation $\hat{A}$ to its adjacency matrix $A$ with LF matrices $X^{n\times K}=[x_{ik}]$ and $Y^{n\times K}=[y_{jk}]$ to yield $\hat{A}=XY^T$ [41], [42], [43], [56]-[58]. To acquire $X$ and $Y$, a cost function is needed to express the difference between $A$ and $\hat{A}$. Thus, such a learning objective is defined based on the Euclidean distance, i.e.,

$$O_{NMF} = \left\| XY^T - A \right\|_F^2, \ s.t. \ X \geq 0, Y \geq 0. \tag{2}$$

where $\|\cdot\|_F$ is the Frobenius norm. $X$ and $Y$ are restrained to be nonnegative for describing nonnegative probabilities standing for each node's community tendency.

Note that $O_{NMF}$ is non-convex in both $X$ and $Y$, making $X$ and $Y$'s global optima intractable. However, $X$ and $Y$'s stationary point can be achieved via an alternative and iterative learning algorithm [59]-[63]. To resolve $X$ and $Y$, the Nonnegative Multiplicative Update (NMU) algorithm is presented by Lee and Seung [49] to optimize $O_{NMF}$ with $X$ and $Y$ as:

$$x_{ik} \leftarrow x_{ik}\left((AY)_{ik}\big/(XY^TY)_{ik}\right), \ y_{jk} \leftarrow y_{jk}\left((A^TX)_{jk}\big/(YX^TX)_{jk}\right). \tag{3}$$

An NMF-based community detector can be established with (3). Note that it acquires the representation of LUN by adopting multiple LF matrices without considering the symmetry, which is its intrinsic property [32].

*C. An SNMF-based Community Detector*

An SNMF-based model adopts a unique LF matrix $X^{n\times K}=[x_{ik}]$ to represent a target LUN's adjacency matrix $A$ only. In particular, its learning objective is given as:

$$O_{SNMF} = \left\| XX^T - A \right\|_F^2, \ s.t. \ X \geq 0, \tag{4}$$

where $X$ can be found via an NMU algorithm, i.e.,

$$x_{ik} \leftarrow x_{ik}\left((AX)_{ik}\big/(XX^TX)_{ik}\right). \tag{5}$$

As discussed in [39], [50], (5) commonly makes an SNMF model suffer from unstable convergence. According to [44], the following learning rule with an adjusted multiplicative term is commonly adopted:

$$x_{ik} \leftarrow x_{ik}\left(0.5 + (AX)_{ik}\big/(2XX^TX)_{ik}\right). \tag{6}$$

Note that an SNMF-based method imposes strong symmetry into its approximation, thus enabling it to represent LUN's symmetry. Moreover, according to [32], it has more general interpretation as a clustering model to explain a universal law in networks: two directly connected nodes probably belong to the same community [32]. However, the unique LF matrix also leads to shrinkage of its LF space, which reduces its representation learning ability to $A$ [32].

III. AN SGNMF-BASED COMMUNITY DETECTOR

*A. Learning Objective and Scheme*

To make an SGNMF model own the symmetry of a target LUN, we put a symmetry regularization term, i.e., $\|X-Y\|_F$, into the learning objective, i.e.,

$$O_{SGNMF} = \frac{1}{2}\left\| XY^T - A \right\|_F^2 + \frac{\alpha}{2}\left\| X - Y \right\|_F^2, \ s.t. \ X \geq 0, Y \geq 0, \tag{7}$$

where $A$ denotes the adjacency matrix. $X$ and $Y$ denote LF matrices. $\alpha>0$ is a coefficient to balance the loss error and the difference between $X$ and $Y$. By (7), $X$ and $Y$ are required to be equal for indirectly representing the symmetry of $A$. On the other hand, this constraint is loose and can be adjusted with $\alpha$: as $\alpha$ increases, $X$ and $Y$ become closer to make an SGNMF model tend to own a target LUN's symmetry, and vice versa. However, LF space is not shrunk with two LF matrices, i.e., $X$ and $Y$. Therefore, (7) also ensures an SGNMF model's high representation ability to $A$.

Meanwhile, to ensure the local invariance of a target LUN, we further incorporate the graph regularization into (7), i.e.,

$$O_{SGNMF} = \frac{1}{2}\left\| XY^T - A \right\|_F^2 + \frac{\alpha}{2}\left\| X - Y \right\|_F^2 + \frac{\lambda}{2}\text{Tr}\left(Y^TLY\right), \ s.t. \ X \geq 0, Y \geq 0, \tag{8}$$

where $\lambda>0$ adjusts graph-regularization effects. $L$ denotes the Laplacian matrix and $L=D-S$. The element in diagonal matrix $D$ is computed as $D_{ii}=\sum_l S_{il}$, and the similarity matrix $S$ measures the closeness among network node pairs. In our context, matrices $A$ and $S$ are numerically equal according to prior work [39]. With the commonly accepted property that $\|A\|_F=\text{Tr}(AA^T)$ as $\text{Tr}(\cdot)$ calculates the trace of a matrix, the objective function in (8) can be reformulated into:

$$O_{SGNMF} = \frac{1}{2}\text{Tr}\left(XY^TYX^T - 2AYX^T + AA^T\right) + \frac{\lambda}{2}\text{Tr}\left(Y^TLY\right) + \frac{\alpha}{2}\text{Tr}\left(XX^T - 2XY^T + YY^T\right), \ s.t. \ X \geq 0, Y \geq 0, \tag{9}$$

Letting $K=[\kappa_{ik}]$ and $\Gamma=[\gamma_{jk}]$ be the Lagrangian multipliers for the nonnegative constraints of $X=[x_{ik}]>0$ and $Y=[y_{jk}]>0$, the Lagrangian function can be obtained as:

$$L_{SGNMF} = \frac{1}{2}\text{Tr}\left(XY^\text{T}YX^\text{T} - 2AYX^\text{T} + AA^\text{T}\right) + \frac{\lambda}{2}\text{Tr}\left(Y^\text{T}LY\right) + \frac{\alpha}{2}\text{Tr}\left(XX^\text{T} - 2XY^\text{T} + YY^\text{T}\right) + \text{Tr}\left(KX^\text{T}\right) + \text{Tr}\left(\Gamma Y^\text{T}\right). \quad (10)$$

The partial derivatives of $L_{SGNMF}$ with $X$ and $Y$ are given as:

$$\partial L_{SGNMF}/\partial X = XY^\text{T}Y - AY + \alpha X - \alpha Y + K, \quad (11)$$

$$\partial L_{SGNMF}/\partial Y = YX^\text{T}X - A^\text{T}X - \alpha X + \alpha Y + \lambda LY + \Gamma. \quad (12)$$

Setting $\partial L_{SGNMF}/\partial X = \partial L_{SGNMF}/\partial Y = 0$, a local minimum of (8) can be gained. Then, based on the KKT conditions that $\kappa_{ik}x_{ik}=0$ and $\gamma_{jk}y_{jk}=0$, we have the solutions of $x_{ik}$ and $y_{jk}$:

$$-(AY + \alpha Y)_{ik} x_{ik} + (XY^\text{T}Y + \alpha X)_{ik} x_{ik} = 0, \quad (13)$$

$$-(A^\text{T}X + \alpha X)_{jk} y_{jk} + (YX^\text{T}X + \alpha Y + \lambda LY)_{jk} y_{jk} = 0. \quad (14)$$

Finally, with (13) and (14), the learning scheme for SGNMF can be achieved as:

$$x_{ik} \leftarrow x_{ik}(AY + \alpha Y)_{ik} / (XY^\text{T}Y + \alpha X)_{ik}, \quad (15a)$$

$$y_{jk} \leftarrow y_{jk}(A^\text{T}X + \alpha X + \lambda AY)_{jk} / (YX^\text{T}X + \alpha Y + \lambda DY)_{jk}. \quad (15b)$$

With (15), an SGNMF-based community detector is obtained. Note that (15) reduces to the learning rule of a Graph-regularized NMF (GNMF)-based method [11], [35] when $\alpha=0$, and also reduces to the learning rule of a standard NMF-based method [49] when $\alpha=\lambda=0$.

IV. PERFORMANCE ANALYSIS

*A. General Settings*

***Evaluation Protocol.*** Following [9], [11], [17], [28], [35], Normalized Mutual Information (NMI) is adopted to evaluate the performance of involved models. NMI measures the similarity between the resulting community assignment and the ground-truth community information. NMI values range from 0 to 1, with larger NMI standing for higher performance of a detection model.

***Datasets.*** As shown in Table II, ten real-world LUNs are used in our empirical studies. They are publicly-available datasets that are widely used in previously related studies.

***Tested Models.*** We compare SGNMF with nine baseline and state-of-the-art community detectors, including NMF [49], GNMF [10], [35], SNMF [44], GSNMF [10], NSED [54], SymNMF [32], DeepWalk [55], LINE [56], HPNMF [28], and SGNMF.

TABLE II
DETAILS OF THE ADOPTED DATASETS

| **Datasets** | *Nodes* | *Edges* | *K* | **Description** |
|---|---|---|---|---|
| Youtube [37] | 11,144 | 36,186 | 40 | Youtube online |
| Flickr [38] | 8,051 | 188,687 | 193 | Flickr social network |
| Friendster [37] | 11,023 | 280,755 | 13 | Friendster online |
| Polblogs [1] | 1,490 | 16,718 | 2 | Blogs about US politics |
| LJ [37] | 7,181 | 253,820 | 30 | LiveJournal online |
| Cornell [41] | 195 | 304 | 5 | Subnetwork of WebKB |
| Dolphins [40] | 62 | 159 | 2 | Dolphin social network |
| Orkut [37] | 11,751 | 270,667 | 5 | Orkut online |
| Amazon [37] | 5,304 | 16,701 | 85 | Amazon product |
| DBLP [37] | 12,547 | 35,250 | 4 | DBLP collaboration |

Among these comparison models, a)-d) are four baseline NMF-based methods which are either asymmetric or absolutely symmetric. We adopt them to show the effects of SGNMF's symmetry-regularization. e)-i) are five state-of-the-art methods for network representation learning and community detection, among which SymNMF also adopts an equality-constraint to model LUN's symmetry, and both DeepWalk and LINE are two deep learning-based models. Note that for DeepWalk and LINE, node representations need to be learned firstly, and the *k*-means algorithm is then adopted to cluster the acquired representations for implementing community detection.

***Settings.*** To achieve objective experiment results, the following commonly accepted settings are adopted:

a) The random initialization of LF matrices commonly affects the performance of an NMF-type model according to [41]. Hence, to eliminate initialization biases, the same randomly generated arrays in the scale of (0, 0.05) are used to initialize NMF-type methods.

b) Since specific hyper-parameters are highly critical for the performance of a tested model, all hyper-parameters of tested models are set with their optimal values. For both DeepWalk and LINE, the experiments are performed by default settings in official toolkits, i.e., setting the walk length at 40 and window size at 5 for DeepWalk; setting walk length at 80, window size at 10, and the initial learning rate at 0.025 for LINE; and setting representation dimension at 64 for both models. For graph regularized models, i.e., GNMF, GSNMF and SGNMF, the graph regularization coefficient, i.e., $\lambda$, is uniformly set at 100 according to [35].

For HPNMF, both of its parameters, i.e., $\lambda$ and $\gamma$, are set at 1. In addition, for SGNMF, we set its symmetry regularization parameter, i.e., $\alpha$ at $2^{-8}$ uniformly on all datasets. Note that NMF, SNMF, NSED and SymNMF do not involve any hyper-parameter.

c) Each tested model's training process terminates if: a) the objective value's difference in two consecutive iterations is smaller than $10^{-1}$; and b) the iteration count reaches 200.

*B. Sensitivity Tests*

Note that SGNMF depends on two hyperparameters, i.e., symmetry regularization coefficient $\alpha$ and graph regularization one $\lambda$. In this part of experiments, we adopt the modularity as the validation indicator to perform SGNMF's hyperparameter sensitivity tests with respect to them. The candidate sets of $\alpha$ and $\lambda$ are respectively set as $\{0, 2^{-10}, 2^{-8}, 2^{-6}, 2^{-4}, 2^{-2}, 1, 2\}$ and $\{0, 10^{-3}, 10^{-2}, 10^{-1}, 1, 10, 10^2, 10^3\}$. The results on D1-D4 are depicted in Fig. 1.

According to Fig. 1, we see that SGNMF's performance is sensitive with $\alpha$ and $\lambda$. For instance, as shown in Fig. 1(a), when $\lambda$ increases in [0, 10] on the YouTube network, SGNMF's modularity increases gradually. However, as $\lambda$ increases over a certain threshold, i.e., ten on the YouTube network, SGNMF's $Q$ value decreases. Considering $\alpha$, the outcomes are similar. For instance, on the Youtube network as shown in Fig. 1(a), when $\lambda=1$, SGNMF's modularity increases as $\alpha$ increases in $[0, 2^{-8}]$, and gradually decreases as $\alpha$ exceeds $2^{-8}$. Similar results are achieved on the other networks: optimal $\alpha$ and $\lambda$ enables the SGNMF-based community detector to achieve high modularity and vice versa. We also note that the optimal $\alpha$ and $\lambda$ are somehow data-dependent. The above results have the following two-fold implications:

a) Both symmetry and graph regularizations have effects on the SGNMF-based community detector's performance;
b) The values of $\alpha$ and $\lambda$ are data-dependent, indicating that the SGNMF's regularization effects should be adjusted to fit the structure differences of different LUNs.

On the other hand, although the optimal $\alpha$ and $\lambda$ are somehow dataset-dependent, a SGNMF-based community detector is able to achieve tolerable accuracy loss as $\alpha$ and $\lambda$ vary in a relatively wide scale. Considering $\alpha$, when it lies in the range of $[2^{-8}, 2^{-4}]$, SGNMF achieves steadily high modularity on most tested networks. In terms of $\lambda$, when it lies in the range of [1, 10], SGNMF generally achieves high modularity as shown in Fig. 1. Hence, for real applications, the hyperparameters $\alpha$ and $\lambda$ can be selected by: a) adopting the modularity-based grid-search as in this section; and b) referring to the empirical values achieved in this study, i.e., $\lambda \in [1, 10]$ and $\alpha \in [2^{-8}, 2^{-4}]$.

*C. Performance Verification*

We compare SGNMF with the baseline and state-of-the-art models to validate its effectiveness. Average NMI values of ten tested models on ten real social networks are recorded in Table III. From the results, we obtain the following findings:

a) **Symmetry regularization enables an original NMF-based community detector to achieve high accuracy effectively.**, SGNMF outperforms GNMF on nine testing cases out of ten in total in NMI as shown in Table III. Hence, the superiority of SGNMF in achieving highly accurate community detection is evident. The main reason for this phenomenon is its incorporation of symmetry regularization into its learning objective, thereby enabling its accurate representation to an LUN's symmetry brought by multiple LF matrices standing for a large LF space.

b) **Symmetry and graph regularizations work compatibly to make the proposed SGNMF outperform its peers in the aspect of detection accuracy.** From Table III, we observe that SGNMF outperforms its peers in community detection accuracy on eight testing cases out of ten in total. Its accuracy gain is evident, which is brought by both symmetry and graph regularization integrated in its learning objective. On the other hand, it clearly outperforms GNMF and SNMF in detection accuracy, indicating that both regularization schemes work compatibly to enable its high accuracy.

*D. The Effects of Symmetry and Graph regularization*

As discussed in Sections I and III, due to the incorporation of both the symmetry and graph regularizations, SGNMF can better represent a target LUN than the NMF and SNMF models do. In Section IV, we have verified such a conclusion. To further analyze the effects of the adopted regularization terms, we compare the performance of a set of SGNMF models with different settings, i.e., M1 with $\alpha=0$ and $\lambda=100$ (which is actually the GNMF model), M2 with $\alpha=2^{-8}$ and $\lambda=0$, and M3 with $\alpha=2^{-8}$ and $\lambda=100$ (which is the proposed SGNMF model). The results are depicted in Fig. 2.

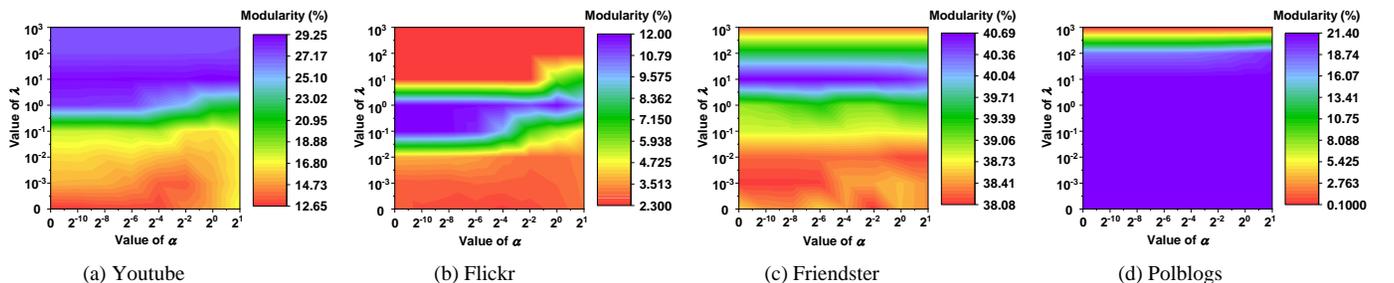

(a) Youtube     (b) Flickr     (c) Friendster     (d) Polblogs

Fig. 1. Sensitivity tests of $\alpha$ and $\lambda$ for SGNMF on D1-D4.

TABLE III
COMMUNITY DETECTION PERFORMANCE (NMI%±STD%) OF TESTED MODELS ON EACH NETWORK, INCLUDING WIN/LOSS COUNTS AND FRIEDMAN TEST

| Models<br>Datasets | NMF | GNMF | SNMF | GSNMF | NSED | SymNMF | DeepWalk | LINE | HPNMF | SGNMF |
|---|---|---|---|---|---|---|---|---|---|---|
| Youtube | 17.26±2.58 | 38.99±2.47 | 16.95±2.10 | 49.45±2.18 | 18.44±1.44 | 18.59±1.85 | 18.43±0.6 | 17.84±1.39 | 49.10±2.38 | **51.79±1.01** |
| Flickr | 0.46±0.13 | 0.32±0.13 | 0.39±$\epsilon$ | 23.36±0.95 | 15.77±1.54 | 0.37±$\epsilon$ | 20.59±0.16 | 20.41±0.19 | **29.42±1.15** | 24.88±1.69 |
| Friendster | 68.08±6.66 | 45.17±2.89 | 67.64±5.53 | 62.64±6.95 | 76.14±5.13 | 78.40±4.49 | 80.37±0.98 | 70.31±1.95 | 80.55±3.55 | **85.24±3.22** |
| Polblogs | 48.41±3.42 | 44.43±8.24 | 44.83±1.35 | 23.49±8.72 | 47.92±3.17 | 45.81±1.36 | 24.87±1.30 | 11.17±2.38 | 44.17±1.31 | **49.45±4.15** |
| LJ | 23.47±3.87 | 49.89±2.41 | 33.78±4.49 | 73.17±3.82 | 19.79±5.19 | 42.22±2.81 | 48.38±2.71 | 42.33±1.59 | 66.82±0.39 | **75.26±2.72** |
| Cornell | 1.77±1.55 | 13.24±5.98 | 0.61±0.18 | 8.42±6.39 | 1.64±1.34 | 0.92±0.01 | 4.29±1.12 | 2.81±0.69 | 11.47±8.86 | **15.22±5.50** |
| Dolphins | 81.43±$\epsilon$* | 65.55±2.46 | 81.43±$\epsilon$ | 53.09±9.26 | 81.43±$\epsilon$ | 81.43±$\epsilon$ | 76.79±0.35 | 64.98±0.71 | 81.43±$\epsilon$ | **88.89±0.00** |
| Orkut | 29.53±2.13 | 51.56±2.62 | 29.39±6.36 | 59.18±4.02 | 31.25±4.93 | 33.75±7.20 | 57.37±1.48 | 42.31±2.36 | 48.13±8.46 | **60.22±0.55** |
| Amazon | 42.21±1.46 | 67.54±1.23 | 45.58±1.72 | 62.29±4.25 | 38.17±1.64 | 47.17±2.07 | 51.98±1.49 | 48.23±1.65 | 61.21±3.93 | **70.95±3.24** |
| DBLP | 7.18±3.22 | **19.18±4.86** | 6.60±2.81 | 9.25±1.69 | 5.67±3.03 | 9.18±2.75 | 11.16±0.76 | 8.92±0.44 | 11.54±1.67 | 16.16±4.32 |
| Win/Loss | 10/0 | 9/1 | 10/0 | 10/0 | 10/0 | 10/0 | 10/0 | 10/0 | 9/1 | — |
| Friedman Ranks | 7.1 | 5.1 | 8 | 4.9 | 7 | 6.2 | 5.1 | 6.9 | 3.5 | **1.2** |

*Note that $\epsilon$ denotes a very small value which is less than $10^{-5}$.

As shown in Fig. 2, in most testing case (nine out of ten in total), symmetry and graph regularizations work compatibly to enable an SGNMF model's high accuracy in community detection. For instance, on D1, M3's NMI is 51.79%, which is about 24.71% (($NMI_{high}$-$NMI_{low}$)/$NMI_{high}$) higher than M1's 38.99%, and about 65.21% higher than M2's 18.02%. On D2-9, similar situations are also encountered as shown in Fig. 2.

The only exception occurs on D10, where M3's NMI is a bit lower than that of M1. Note that as discussed before, with the symmetry regularization, *X* and *Y* are forced to be equal to some extent, which will inevitably lead to reduction of the LF space. Hence, the exceptional results on D10 indicate that on a small portion of LUNs, keeping a large LF space is more important than illustrating its symmetry.

*E. Summary*

Based on the experiment results, we conclude that: a) Both symmetry and graph regularizations work compatibly to enable SGNMF's accurate representation to an LUN; b) The symmetry regularization effects should be carefully adjusted to balance SGNMF's LF space and structure representation; and c) An SGNMF-based community detector outperforms both baseline and state-of-the-art community detectors in terms of detection accuracy significantly.

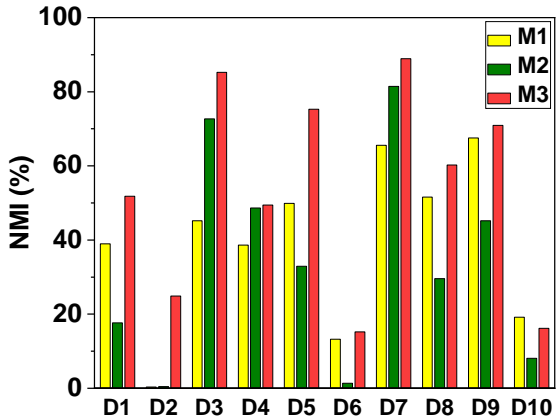

Fig. 2. Performance comparison among a set of SGNMF models with different settings, i.e., M1 with $\alpha$=0 and $\lambda$=100, M2 with $\alpha$=$2^{-8}$ and $\lambda$=0, and M3 with $\alpha$=$2^{-8}$ and $\lambda$=100. D1-D10 denote networks given in Table II in order.

## V. CONCLUSION

An LUN's symmetry is its intrinsically structural feature. When performing community detection on an LUN, an SNMF model mostly adopts a unique LF matrix for accurately demonstrate its symmetry, which leads to a reduced LF space that impairs its representation learning ability. To tackle this problem, this paper presents a novel SGNMF model that adopts symmetry and graph regularizations to accurately represent an LUN. Experiment results on ten real-world LUNs show that an SGNMF-based detector achieves higher community detection accuracy than baseline and state-of-the-art methods.

The future work plans to tackle the issue of hyper-parameter adaptation via evolutionary computation techniques such as a particle swarm optimization (PSO) algorithm [61]-[63].